\documentclass[showkeys,aps,graphicx,twocolumn]{revtex4}
\usepackage{amssymb}
\usepackage{amsmath}
\usepackage{graphicx}
\usepackage{array}
\usepackage{subfigure}
\usepackage{multirow}
\usepackage[colorlinks,
linkcolor=blue,
anchorcolor=blue,
citecolor=blue,
urlcolor=blue
]{hyperref}

\begin{document}

\title{Error-detected three-photon hyperparallel
	Toffoli gate with state-selective reflection}

\author{Yi-Ming Wu, Gang Fan and Fang-Fang Du\footnote{Corresponding author:
Duff@nuc.edu.cn} }
\affiliation{  Science and Technology on Electronic Test
	and Measurement Laboratory, North University of China, Taiyuan 030051, China }
\date{\today }

\begin{abstract}
We present an error-detected hyperparallel Toffoli (hyper-Toffoli) gate for a three-photon system based on the interface between polarized photon and cavity-nitrogen-vacancy(NV) center system. This hyper-Toffoli gate can be used to perform double Toffoli gate operations simultaneously on both the polarization and spatial-mode degrees of freedom (DoFs) of a three-photon system with a low decoherence, shorten operation time, and less quantum resources required, in compared with those on two independent three-photon systems in one DoF only.
As the imperfect cavity-NV-center interactions are transformed into the detectable failures rather than infidelity
based on the heralding mechanism of detectors,
a near-unit fidelity of the quantum hyper-Toffoli gate can be implemented.
By recycling the procedures, the efficiency of our protocol for the hyper-Toffoli gate is improved further. Meanwhile,
the evaluation of gate performance with achieved experiment parameters shows that it is feasible with current experimental technology and provides a promising building block for quantum compute.\\

\end{abstract}

\maketitle

\section{Introduction}      \label{sec1}

Compared to the best-known classical information processing,
quantum information processing (QIP) \cite{Quantum2002}, adhered to the principles of quantum mechanics, has great advantages in secure communication \cite{Error2019, Quantum2020, Generic2020, Measurement2021,Quantum2021,Wang2021}, algorithm \cite{Polynomial-Time1999,Quantum1997,Cavity2018}, and machine learning \cite{Entanglement2015,Quantum2018,Experimental2020, Machine2021}, due to its great security and super-fast computing speed.
Basic quantum logic gates are the building block of quantum computing, so they are regarded as an important unit of QIP.  However, they still face the challenge of large-scale integration in practice.
Therefore, to simplify the constantly growing complexity of qubits and decrease the overheads of resources, multi-qubit gates become the central ingredients in scalable QIP.  Generally,
the simplest universal multi-qubit gates is Toffoli or Fredkin gate \cite{Optimal2020,Low-cost2020}.
To construct a $n$-qubit Toffoli gate, $[(4^n-3n-1)/4]$
two-qubit gates by Shende \emph{et al.} \cite{Minimal2004} can be required. Ralph \emph{et al.} introduced qudit to model Toffoli gate with polynomial two-qubit gates \cite{Efficient2007}, where the qudit takes advantage of  ancillary higher-dimensional spaces.
Based on multi-level physical systems, Radu \emph{et al.} \cite{Generalized2009} presented the generalized $ n $-qubit Toffoli gate using $ n+1 $ subsystems.
Yu \emph{et al.} deemed five two-qubit gates as optimal for implementing the three-qubit Toffoli gate in theory \cite{Five2013}.  Fiur\'{a}\v{s}ek \emph{et al.}  \cite{Linear2006} proposed a
linear-optics-based quantum Toffoli gate. Soeken \emph{et al.} \cite{Quantum2013} investigated the relationships between Pauli matrices and quantum logical gates for simplifying the quantum circuits.
Recently,
some physical architectures, including superconducting circuits \cite{Implementation2012,Single-step2020,Machine-learning-based2020}, nitrogen-vacancy (NV)  centers \cite{Compact2013}, quantum dots (QDs) \cite{Scalable2014,Universal2014O,Heralded2021}, and trapped ions, \cite{Realization2009} have been proposed to implement extensive
Toffoli gates in experiment and theory.

Less resources required from quantum gates is crucial
in quantum computation, originating from two
strategies. One is to exploit a system with additional qudit \cite{Fast2013, Simplifying2009, Scalable2020}, the other one is to encode the quantum information in multiple degrees of freedom (DoFs) of a system in universal computational tasks, i.e., hyperparallel quantum gates \cite{ Universal2015,Hyperparallel2016,Robust2019,Heralded2021O,Universal2014L,Realization2021,Wei:16}  simultaneously operating more than one independent operations. The  hyper-parallel quantum computing can  simplify the
quantum circuit, improve the information-processing speed,  reduce the quantum
resource consumption, and suppress the photonic dissipation noise.
Recently,
implementing hyper-parallel photon-based controlled-NOT (CNOT) gate \cite{Universal2015,Hyperparallel2016,Robust2019,Heralded2021O}, hyper-parallel photon-matter-based universal CNOT gate \cite{Universal2014L}, and hyper-parallel photon-based Toffoli gate \cite{Wei:16} have been proposed. Ru \emph{et al.} \cite{Realization2021} have realized a deterministic quantum Toffoli gate on one photon in orbital-angular-momentum and polarization DoFs.
Therefore, enquiry of the hyper-parallel multi-qubit Toffoli gate will be provided with great significance in scalable hyper-parallel QIP.

Up to now,  solid-state-spin systems play a
promising platform for the realization of quantum computing,  as solid-state nature combined with nanofabrication
techniques provide a useful way to incorporate the electronic
spins into optical microcavities \cite{Room-temperature2006,Gigahertz2009,Deterministric2010,Experimental2015}.
One appealing type of
solid-state-spin system is the electron spin in a NV center \cite{High-filelity2011,Quantum2007,Room2010},
which
not only provides convenient ways of optical
initialization, single-qubit manipulation, and high-fidelity readout at room
temperature, but also
processes a milliseconds coherence time of the electron spin in
NV center, resorting to
spin echo techniques \cite{Decoherence-protected2012,Room2014}.
In recent years, the NV centers have numerous applications, such as distributed quantum computing \cite{Experimental2014},  the hybrid quantum gates acting on the electron spin and the nearby nuclear spin \cite{Quantum2010},
geometric single-qubit gates \cite{Entangled2015}, CNOT gate \cite{Heralded2013,Robust2019}, and universal gates on NV center electronic \cite{Universal2015} or photonic qubits \cite{Hyperentanglement2013,Universal2014P} , and so on. Moreover, NV-center-emission-based entanglement \cite{Universal2013} and hyperentanglement
purification \cite{Efficient2019} were also proposed recently.

In this paper, we propose an error-detected
hyper-parallel Toffoli gate
for a three-photon system assisted by the state-selective reflection of one-sided cavity-NV-center system.
Our scheme has some characters. First,
this hyper-Toffoli gate is equivalent to perform double Toffoli gate operations simultaneously on both the polarization and spatial DoFs of a three-photon system with a low decoherence, short operation time, and less quantum resources required, in compared with those on two independent three-photon systems in one DoF only.
Second, the fidelity of the quantum hyper-Toffoli gate is unity in principle,  as the inevitable imperfect performances can be detected by single-photon detectors. Third, the strong
coupling limitations can be avoided, reducing the experimental conditions. Fourth, the success of the error-detected scheme can be heralded by the single-photon
detectors. Fifth, the efficiency of the scheme can be further improved by repeating the operation processes when
the detectors are clicked.

\section{ Hyper-paralleled Quantum Toffoli Gate}    \label{sec2}

\begin{figure}[tbh!]
	\centering
	\includegraphics[width=0.7\linewidth]{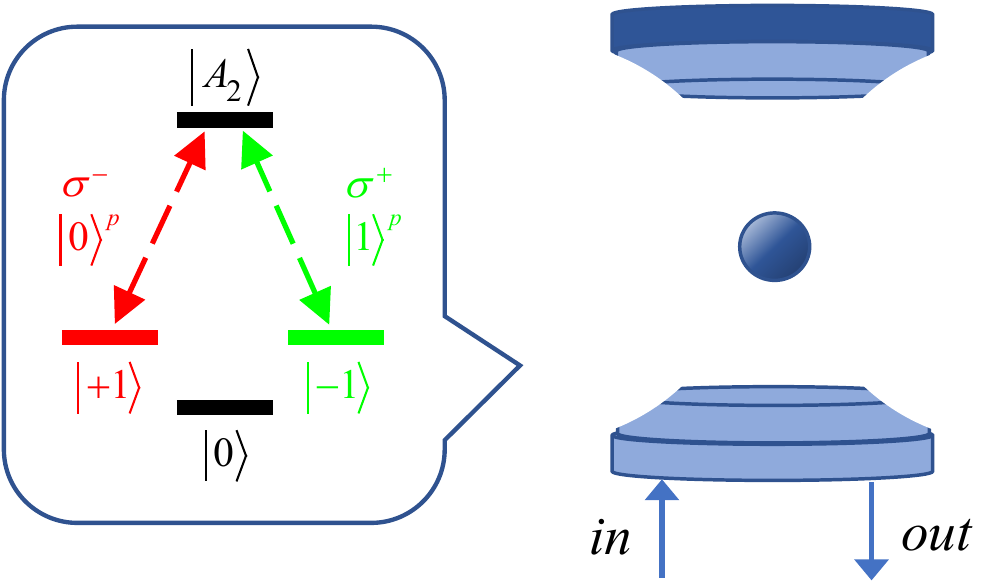}
	\caption{ A cavity-NV center system consists of the negatively
		charged NV center confined in an optical one-sided cavity.  The optical transitions from the spin-ground
		states $|\pm1\rangle$ to the excited state $|A_{2}\rangle$ are coupled by the $|\sigma_{\mp}\rangle$ circularly polarized photons. }
	\label{fig1}
\end{figure}

The cavity-NV center system is originated from
a diamond NV center that is coupled to an optical one-sided cavity,
as shown in Fig. \ref{fig1}. The bottom mirror of the optical cavity
is partially reflection and the top one is 100\% reflection.
The negatively charged NV center in the diamond lattice is
composed by a substitutional nitrogen atom, an adjacent vacancy and six electrons.
Here, the spin-ground state of the negatively charged NV center owing to spin-spin interaction is split into $|0\rangle (m_{s}=0)$ and $|\pm1\rangle (m_{s}=\pm1)$ with a 2.87GHz zero-field splitting.
$|A_{2}\rangle=(|E_{-}\rangle|+1\rangle+|E_{+}\rangle|-1\rangle)/\sqrt{2}$ is one of
the six excited states of the negatively charged NV center \cite{ Quantum2010,Entangling2011}, taking into account spin-orbit and spin-spin interactions at the same time, and is robust due to stable symmetric property.
$|E_{\pm}\rangle| (J_{s}=\pm1)$ is orbit states of the cavity-NV center system.
The optical transition $|+1\rangle \leftrightarrow |A_{2}\rangle$
($|-1\rangle \leftrightarrow |A_{2}\rangle$) is driven by the absorption or emission of
a left-(right-) circularly polarized photon $|R\rangle$ ($|L\rangle$), which can be improved
by the NV center trapped in a frequency-degenerate two-mode cavity.
The above polarized states can be rewrote as $|R\rangle = |0\rangle^p$, $|L\rangle = |1\rangle^p$ in the following presentation.
Under the rotating wave approximation, the Hamiltonian of the cavity-NV-center system can be written as \cite{BOOK1997}
\begin{eqnarray} \label{eq0}
	H= \omega_{d}\hat{\sigma}_{+}\hat{\sigma}_{-}+\omega_{c}\hat{a}^{\dagger}\hat{a}+ig(\hat{\sigma}_{+}\hat{a}-\hat{a}^{\dagger}\hat{\sigma}_{-}),
\end{eqnarray}
where $ \omega_{d} $ and $ \omega_{c} $ denote the frequencies of the negatively charged NV center and the cavity field, respectively.
$ \hat{\sigma}_{+} $ and $ \hat{\sigma}_{-} $ are the lifting and lowering operators of the negatively charged NV center.
$ \hat{a} $ and $\hat{a}^{\dagger}$ are the annihilation and creation operators of the cavity filed.
$g$ is the coupling strength between the NV center and the one-sided cavity.

The Heisenberg-Langevin equations about the operators $\hat{a}$ and
$\hat{\sigma}_{-}$, and the input-output relationship of cavity filed could be described as \cite{Deterministic2008}
\begin{eqnarray}\label{eq1}
	\frac{d\hat{a}}{dt} &=& -\left[i\left(\omega_{c}-\omega_{p}\right)+\frac{\kappa}{2}\right]\hat{a}-g\hat{\sigma}_{-}\nonumber\\
	&&-\sqrt{\kappa}\hat{a}_{in}+\hat{H},\nonumber\\
	\frac{d\hat{\sigma}_{-}}{dt} &=& -\left[i\left(\omega_{d}-\omega_{p}\right)+\frac{\gamma}{2}\right]\hat{\sigma}_{-}-g\hat{\sigma}_{z}\hat{a}\nonumber\\
	&&+\sqrt{\gamma}\hat{\sigma}_{z}\hat{b}_{in}+\hat{G},\nonumber\\
	\hat{a}_{out} &=& \hat{a}_{in}+\sqrt{\kappa}\hat{a}.
\end{eqnarray}
Here, $\omega_{p}$ represents the frequency of the input photon. $ \hat{\sigma}_{z}=\hat{\sigma}_{+}\hat{\sigma}_{-}-\hat{\sigma}_{-}\hat{\sigma}_{+} $ is the population operator. $\kappa$ and $\gamma$ are the decay rate of the cavity field and the electron-spin state in the negatively charged NV center, respectively. $\hat{H}$ and $\hat{G}$ are the noise operators. $\hat{a}_{in}$ and $\hat{a}_{out}$ are the input and output vacuum filed operators, respectively. Under the weak excitation $\left(\langle\hat{\sigma}_{z}\rangle\approx-1\right)$, the reflection coefficient $r_{1}$ of the input photon interacting with the one-sided cavity-NV center system can be described as \cite{Quantum-information2009}
\begin{eqnarray}\label{eq2}
 r_{1}=\frac{\left[i(\omega_{c}-\omega_{p})-\frac{\kappa}{2}\right]\left[i(\omega_{t}-\omega_{p})+\frac{\gamma}{2}\right]+g^2}{\left[i(\omega_{c}-\omega_{p})+\frac{\kappa}{2}\right]\left[i(\omega_{t}-\omega_{p})+\frac{\gamma}{2}\right]+g^2}.
\end{eqnarray}
When the input photon encounters a cold cavity, that is, the coupling strength $g$ = 0, the reflection coefficient $r_{0}$ can be changed into
\begin{eqnarray}\label{eq3}
r_{0}= \frac{i(\omega_{c}-\omega_{p})-\frac{\kappa}{2}}{i(\omega_{c}-\omega_{p})+\frac{\kappa}{2}}.
\end{eqnarray}
Thus in the realistic condition, the state-selective reflection of circularly polarized photon interacting with one-sided cavity-NV center
system can be summarized as
\begin{eqnarray} \label{eq4}
	&&|0\rangle^p|+1\rangle\rightarrow r_{1}|0\rangle^p|+1\rangle,   \nonumber\\ &&|1\rangle^p|+1\rangle\rightarrow r_{0}|1\rangle^p|+1\rangle,\nonumber\\
	&&|0\rangle^p|-1\rangle\rightarrow r_{0}|0\rangle^p|-1\rangle,\nonumber\\ &&|1\rangle^p|-1\rangle\rightarrow r_{1}|1\rangle^p|-1\rangle.
\end{eqnarray}

In this section, we will construct a near-unit fidelity and heralded  hyper-Toffoli gate based on the above state-selective reflection in
Eq. (\ref{eq4}). The Toffoli gate completes the function that the target qubit is flipped only when the two control qubits are both in the state $|1\rangle$, there is no operation on the target qubit when the two control qubits is in the state $|0\rangle$.
The hyper-Toffoli gate simultaneously performs
the Toffoli gate operations on the spatial-polarization DoFs of the three-photon system.

\begin{figure}[tbh!]
	\centering
	\includegraphics[width=0.9\linewidth]{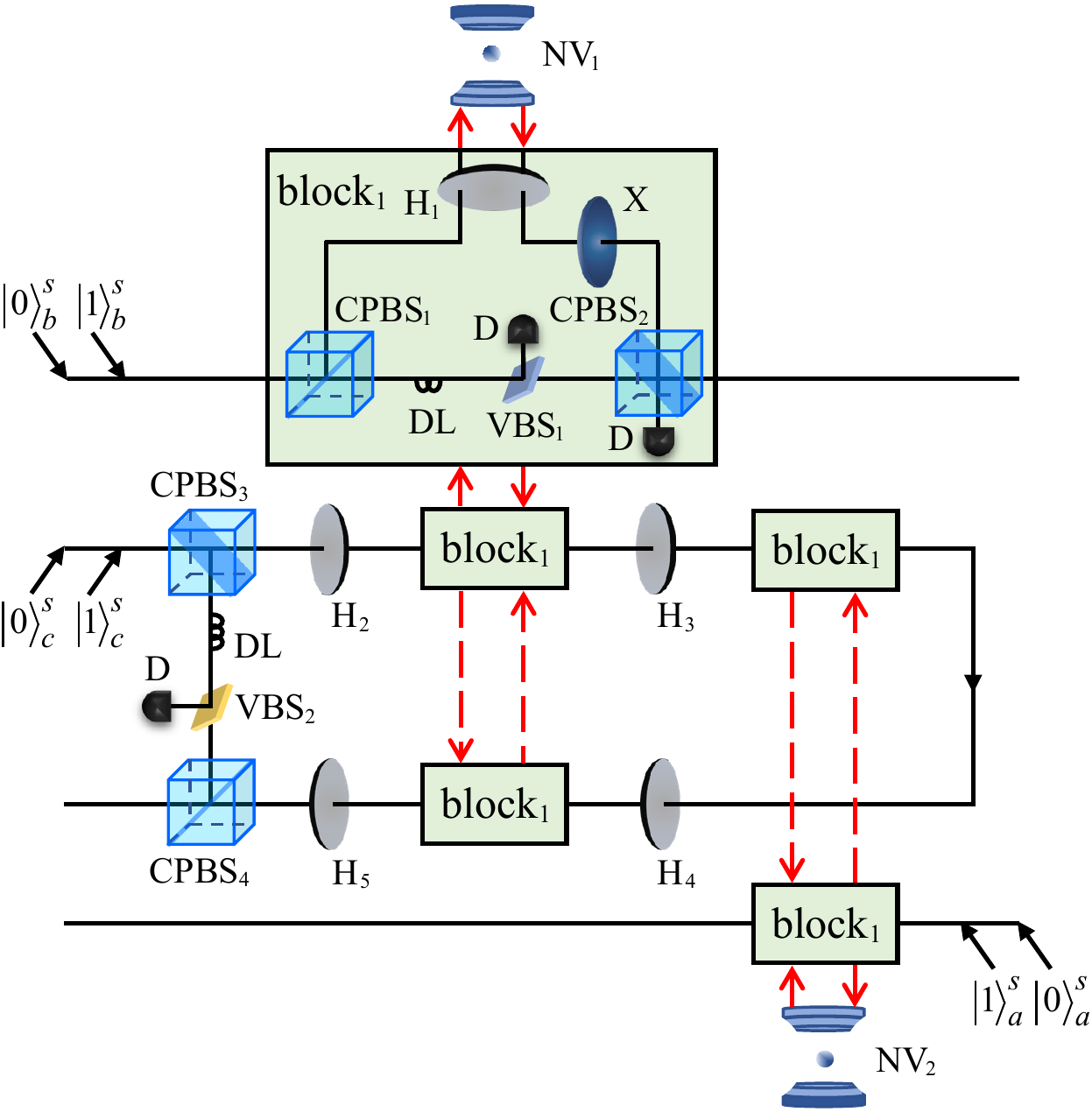}
	\caption{Schematic diagram for the hyper-parallel Toffoli gate for a three-photon system in polarized DoF assisted by two NV-cavity systems. CPBS$_{j}$ ($j$ = 1, 2, 3, 4) represents a circularly polarized beam splitter, which completes the reflection of state $|0\rangle^p$ and the transmission of state $|1\rangle^p$. H$_{k}$ ($k$ = 1, 2, 3, 4, 5) is a half-wave plate inclined at $22.5^{\circ}$, which performs the polarized Hadamard operation, i.e., $|0\rangle^p\rightarrow(|0\rangle^p+|1\rangle^p)/\sqrt{2},  |1\rangle^p\rightarrow(|0\rangle^p-|1\rangle^p)/\sqrt{2}$. X performs a  qubit-flip operation on the polarized DoF of a photon $[|0\rangle^p\leftrightarrow|1\rangle^p]$ by tilting the half-wave plate at $45^{\circ}$. VBS$_{1}$ is an adjustable beam splitter with transmission coefficient $(r_{1}-r_{0})/2$ and reflection coefficient $\sqrt{1-[(r_{1}-r_{0})/2]^2}$. Similarly, the transmission and reflection coefficient of VBS$_{2}$ are $[(r_{1}-r_{0})/2]^3$ and $\sqrt{1-[(r_{1}-r_{0})/2]^6}$, respectively.	D represents a single-photon detector. DL is a delay line, which makes the two wave packets confluent at the same time.}
	\label{fig2}
\end{figure}

\begin{figure}[htp]
	\includegraphics[width=1\linewidth]{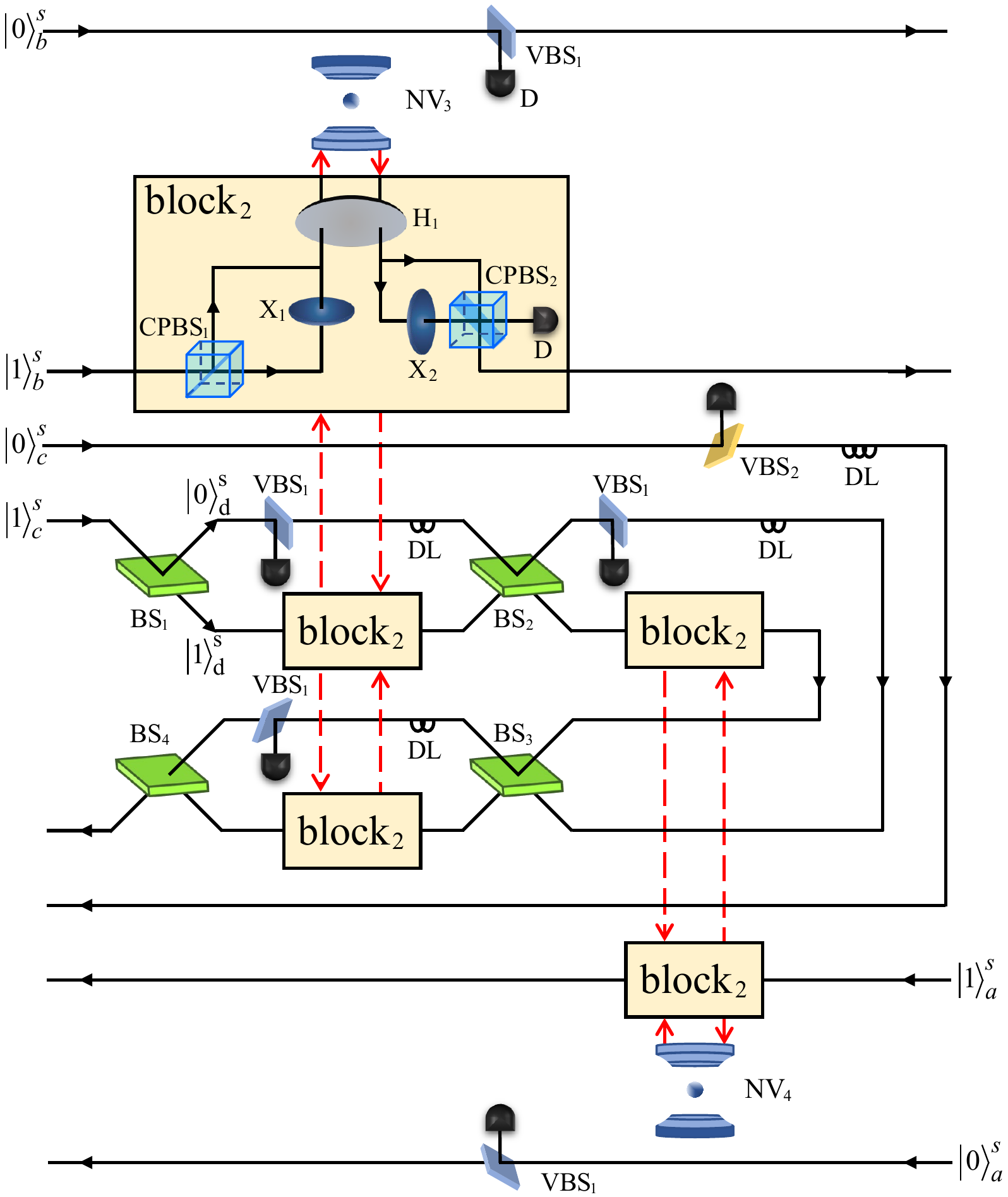}
	\caption{Schematic diagram for the hyper-parallel Toffoli gate for a three-photon system in spatial DoF assisted by two NV-microcavity systems. BS$_{1}$ and BS$_{4}$ denote 50:50 beam splitter which completes the Hadamard operation of the photon in spatial DoF, i.e.,  $|1\rangle_{c}^s\leftrightarrow(|0\rangle_{d}^s+|1\rangle_{d}^s)/\sqrt{2}$.  BS$_{2}$ and BS$_{3}$ complete the operation, i.e., $|0\rangle_{d}^s\leftrightarrow(|0\rangle_{d}^s+|1\rangle_{d}^s)/\sqrt{2}, |1\rangle_{d}^s\leftrightarrow(|0\rangle_{d}^s-|1\rangle_{d}^s)/\sqrt{2} $. }
	\label{fig3}
\end{figure}

The quantum circuit of the error-detected hyper-Toffoli gate can be divided into
two parts. One is used to perform the Toffoli gate
operations on the polarized DoF of a three-photon
system (shown in Fig. \ref{fig2}) and the other is on spatial DoF
(shown in Fig. \ref{fig3}).
Here, the building block$_{1}$ in the green frame interacts with the external cavity-NV$_{i}$ center systems, ($i$=1, 2). Suppose that the initial states of NV$_{1}$ and NV$_{2}$ are both $|-1\rangle$ and the initial polarized and spatial states of three photons are shown as follows
\begin{eqnarray} \label{eq5}
	|\phi\rangle_{a}^{p}=&\alpha_{1}|0\rangle_{a}^p+\alpha_{2}|1\rangle_{a}^p, \quad
	|\phi\rangle_{a}^{s}&=\delta_{1}|0\rangle_{a}^s+\delta_{2}|1\rangle_{a}^s, \nonumber\\
	|\phi\rangle_{b}^{p}=&\beta_{1}|0\rangle_{b}^p+\beta_{2}|1\rangle_{b}^p,  \quad\;
	|\phi\rangle_{b}^{s}&=\epsilon_{1}|0\rangle_{b}^s+\epsilon_{2}|1\rangle_{b}^s, \nonumber\\
	|\phi\rangle_{c}^{p}=&\gamma_{1}|0\rangle_{c}^p+\gamma_{2}|1\rangle_{c}^p, \quad\
	|\phi\rangle_{c}^{s}&=\zeta_{1}|0\rangle_{c}^s+\zeta_{2}|1\rangle_{c}^s.
\end{eqnarray}
The coefficients of the same type satisfy the normalized relationship, i.e. $\lvert Y_{1} \rvert^2+\lvert Y_{2} \rvert^2 = 1$ ($Y=\alpha$, $\beta$, $\gamma$, $\delta$, $\epsilon$, $\zeta$ ). We will present our proposals in order of function.

In the first part shown in Fig. \ref{fig2}, the Toffoli gate operations are performed on the polarized  DoF of three-photon
system $abc$, unaffecting the spatial states of the three photons.
First, let the photons $a$ and
$b$ simultaneously enter into path $|0\rangle_{a}^s$ (or $|1\rangle_{a}^s$) and $|0\rangle_{b}^s$ (or $|1\rangle_{b}^s$) from opposite directions, respectively,  i.e., photon $a$ interacts with the block$_{1}$-NV$_{2}$, meanwhile, photon $b$ interacts with the block$_{1}$-NV$_{1}$.
In particular, before the two photons $a$ and
$b$ pass through block$_{1}$-NV$_{i}$ (i=1, 2), the Hadamard operations H$_{e}$, i.e., $|+1\rangle\leftrightarrow(|+1\rangle+|-1\rangle)/\sqrt{2}, |-1\rangle\leftrightarrow(|+1\rangle-|-1\rangle)/\sqrt{2}$, are performed on the electron-spin state in NV$_{i}$ (i=1, 2) center by using a $\pi /2$ femtosecond-level optical pulse \cite{Picosecond2018}.
In the block$_{1}$, the polarized state $|0\rangle^p$ is reflected by the first circularly polarizing beam splitter (CPBS), then it interacts with H$_{1}$, NV$_{i}$, H$_{1}$ and X in sequence, while the polarized state $|1\rangle^p$ is transmitted to a delay line (DL) and an adjustable beam splitter (VBS). After that, the two optical wave packets $|1\rangle^p$ and $|0\rangle^p$ converge on another CPBS$_{2}$.
When all detectors Ds in the block$_{1}$ are no response, the whole system is transmitted from $|\phi_{0}\rangle$ into $|\phi_{1}\rangle$. Here,
\begin{eqnarray} \label{eq6}
	\!\!\!\!\!\!\!\!	|\phi_{0}\rangle&=&|\phi\rangle_{a}^{p}\otimes|\phi\rangle_{a}^{s}\otimes|\phi\rangle_{b}^{p}\otimes|\phi\rangle_{b}^{s}\otimes|\phi\rangle_{c}^{p}\otimes|\phi\rangle_{c}^{s}\nonumber\\
	&&\otimes|-1\rangle_{1}|-1\rangle_{2},\nonumber\\
	\!\!\!\!\!\!\!\!|\phi_{1}\rangle&=&\frac{r_{1}-r_{0}}{2\sqrt{2}}[|+1\rangle_{2}(\alpha_{1}|0\rangle_{a}^p+\alpha_{2}|1\rangle_{a}^p)\nonumber\\
	&&+|-1\rangle_{2}(\alpha_{1}|0\rangle_{a}^p-\alpha_{2}|1\rangle_{a}^p)]\nonumber\\
	&&\frac{r_{1}-r_{0}}{2\sqrt{2}}[|+1\rangle_{1}(\beta_{1}|0\rangle_{b}^p+\beta_{2}|1\rangle_{b}^p)\nonumber\\
	&&+|-1\rangle_{1}(\beta_{1}|0\rangle_{b}^p-\beta_{2}|1\rangle_{b}^p)]\nonumber\\
	&&\otimes|\phi\rangle_{a}^{s}|\phi\rangle_{b}^{s}|\phi\rangle_{c}^{p}|\phi\rangle_{c}^{s}.
\end{eqnarray}
From Eq. (\ref{eq6}),  the function of block$_{1}$-NV$_{i}$ is equal to performing the controlled-phase-flip (CPF) gate operation on the polarized
state $|0\rangle^{p}$ of photon $a$ or $b$ when the electron-spin state are $|-1\rangle$ in NV$_{i}$ ($i$=1, 2) centers without affecting its spatial state.
When the single-photon detector D in the block$_{1}$ on the spatial mode $|0\rangle_{a}^s$ ($|1\rangle_{a}^s$, $|0\rangle_{b}^s$, or $|1\rangle_{b}^s$) responses,  it means that
the errors originated from the imperfect photon-spin
interaction are filtered by CPBS$_{2}$ and heralded by the D in the
block$_{1}$.
That is,  the
block$_{1}$ has an error-detected function.
After photon $a$ passes through block$_{1}$-NV$_{2}$ and photon $b$ passes through block$_{1}$-NV$_{1}$, the electron-spin state in NV$_{i}$ ($i$=1, 2) centers perform
the Hadamard operation H$_{e}$ again. The quantum state $|\phi_{1}\rangle$ is changed into $|\phi_{1'}\rangle$
\begin{eqnarray} \label{eq7}
	|\phi_{1'}\rangle&=&(\frac{r_{1}-r_{0}}{2})^2\left(\alpha_{1}|0\rangle_{a}^p|+1\rangle_{2}+\alpha_{2}|1\rangle_{a}^p|-1\rangle_{2}\right)\nonumber\\
	&&\left(\beta_{1}|0\rangle_{b}^p|+1\rangle_{1}+\beta_{2}|1\rangle_{b}^p|-1\rangle_{1}\right)\nonumber\\
	&&\otimes|\phi\rangle_{a}^{s}|\phi\rangle_{b}^{s}|\phi\rangle_{c}^{p}|\phi\rangle_{c}^{s}.
\end{eqnarray}

Second, the photon $c$  is put on the quantum circuit. After the CPBS$_{3}$, the polarized
state $|1\rangle_{c}^p$ passes through H$_{2}$, block$_{1}$-NV$_{1}$, H$_{3}$, block$_{1}$-NV$_{2}$,  H$_{4}$, block$_{1}$-NV$_{1}$ and H$_{5}$ in sequence, meanwhile, the polarized
state $|0\rangle_{c}^p$ passes through DL and VBS$_{2}$ with transmission coefficient $[(r_{1}-r_{0})/2]^3$ and reflection coefficient $\sqrt{1-[(r_{1}-r_{0})/2]^6}$. Then the two wave packets cross at another CPBS$_{4}$ simultaneously by DL. When all detectors Ds in two block$_{1}$s do not response, the above operations change the quantum state $|\phi_{1'}\rangle$ into
\begin{eqnarray}   \label{eq8} |\phi_{2}\rangle&=&(\frac{r_{1}-r_{0}}{2})^5[\alpha_{1}|0\rangle_{a}^p|+1\rangle_{2}(\beta_{1}\gamma_{1}|0\rangle_{b}^p|0\rangle_{c}^p|+1\rangle_{1}\nonumber\\
	&&+\beta_{1}\gamma_{2}|0\rangle_{b}^p|1\rangle_{c}^p|+1\rangle_{1}
	+\beta_{2}\gamma_{1}|1\rangle_{b}^p|0\rangle_{c}^p|-1\rangle_{1}\nonumber\\
	&&+\beta_{2}\gamma_{2}|1\rangle_{b}^p|1\rangle_{c}^p|-1\rangle_{1})\nonumber\\
	&&+\alpha_{2}|1\rangle_{a}^p|-1\rangle_{2}(\beta_{1}\gamma_{1}|0\rangle_{b}^p|0\rangle_{c}^p|+1\rangle_{1}\nonumber\\
	&&+\beta_{1}\gamma_{2}|0\rangle_{b}^p|1\rangle_{c}^p|+1\rangle_{1}
	+\beta_{2}\gamma_{1}|1\rangle_{b}^p|0\rangle_{c}^p|-1\rangle_{1}\nonumber\\
	&&-\beta_{2}\gamma_{2}|1\rangle_{b}^p|1\rangle_{c}^p|-1\rangle_{1})]\otimes|\phi\rangle_{a}^{s}|\phi\rangle_{b}^{s}|\phi\rangle_{c}^{s}.
\end{eqnarray}

Third, we apply Hadamard operations H$_{e}$ on the two electron-spin states in NV$_{1}$ and NV$_{2}$, respectively,  and  measure them with the basis  $\left\{|+1\rangle,|-1\rangle\right\}$. If the measurement results of the two-electron-spin states are $|+1\rangle_{1}$ and $|+1\rangle_{2}$, The state of the system
is changed from $|\phi_{2}\rangle$  to
\begin{eqnarray}   \label{eq9} |\phi_{3}\rangle&=&(\frac{r_{1}-r_{0}}{2})^5[(\alpha_{1}\beta_{1}|0\rangle_{a}^p|0\rangle_{b}^p+\alpha_{1}\beta_{2}|0\rangle_{a}^p|1\rangle_{b}^p\nonumber\\
	&&+\alpha_{2}\beta_{1}|1\rangle_{a}^p|0\rangle_{b}^p)(\gamma_{1}|0\rangle_{c}^p+\gamma_{2}|1\rangle_{c}^p)\nonumber\\
	&&+(\alpha_{2}\beta_{2}|1\rangle_{a}^p|1\rangle_{b}^p)(\gamma_{1}|0\rangle_{c}^p-\gamma_{2}|1\rangle_{c}^p)]\nonumber\\
	&&
	\otimes|\phi\rangle_{a}^{s}|\phi\rangle_{b}^{s}|\phi\rangle_{c}^{s},
\end{eqnarray}
which is the result of the controlled-controlled-phase-flip (C$^2$PF) gate on the polarized DoF of a three-photon
system without affecting its spatial state. That is, when the two electron-spin states in NV$_{1}$ and NV$_{2}$ centers, and the
polarized states of two photons $a$ and
$b$  are $|+1\rangle_{1}|+1\rangle_{2}$ and $|1\rangle_{a}^p|1\rangle_{b}^p$, the CPF gate operation on the polarized
state of photon $c$.
However, the measurement results of the two electron-spin states are other possible outcomes, the corresponding feed-forward single-qubit operations are shown in Table \ref{tab1}. Here, $\sigma_{z}^p=|0\rangle^p\langle 0|-|1\rangle^p\langle 1|$ completes the phase-flip operation on polarized state of the photon, and $I$ means keeping its original state. Therefore, the success probability of a C$^2$PF gate on the polarized DoF of a three-photon
system is 100\% in principle.
Finally, before and after the C$^2$PF gate,  Hadamard operation performed on the polarized DoF of the target qubit $c$, i.e., $U^p_{Toffoli} =
(H^p \otimes I_{4})U^p_{C^2PF}(I_{4} \otimes H^p)$, the Toffoli gate on the polarized DoF of the three-photon system is finished without impacting on the spatial DoF of the three-photon system. Here,
\begin{eqnarray}   \label{eq10} |\phi_{4}\rangle&=&(\frac{r_{1}-r_{0}}{2})^5[(\alpha_{1}\beta_{1}|0\rangle_{a}^p|0\rangle_{b}^p+\alpha_{1}\beta_{2}|0\rangle_{a}^p|1\rangle_{b}^p\nonumber\\
	&&+\alpha_{2}\beta_{1}|1\rangle_{a}^p|0\rangle_{b}^p)(\gamma_{1}|0\rangle_{c}^p+\gamma_{2}|1\rangle_{c}^p)\nonumber\\
	&&+(\alpha_{2}\beta_{2}|1\rangle_{a}^p|1\rangle_{b}^p)(\gamma_{1}|1\rangle_{c}^p+\gamma_{2}|0\rangle_{c}^p)]\nonumber\\
	&&\otimes|\phi\rangle_{a}^{s}|\phi\rangle_{b}^{s}|\phi\rangle_{c}^{s}.
\end{eqnarray}

\begin{table} [hpt]
	\tabcolsep 0pt \caption{The measurement results of the two-electron-spin states in NV$_{1}$ and NV$_{2}$, and the corresponding feed-forward single-qubit operations on polarized DoF of three photons.} 
	\begin{center}
		\begin{tabular}{cc}
			\hline
			measurement results \ \ \ \ \
			&single-qubit operations \\\hline	
			$|+1\rangle_{1}|+1\rangle_{2} $ \ \ \ \ \
			&$I_{a}\otimes I_{b}\otimes I_{c}$ \\
			\!\!$|+1\rangle_{1}|-1\rangle_{2} $\ \ \ \ \
			& $(\sigma_{z}^p)_{a}\otimes I_{b}\otimes I_{c}$ \\
			\!\!$|-1\rangle_{1}|+1\rangle_{2}$\ \ \ \ \
			& $I_{a}\otimes (\sigma_{z}^p)_{b}\otimes I_{c}$ \\
			$|-1\rangle_{1}|-1\rangle_{2}$ \ \ \ \ \
			&$(\sigma_{z}^p)_{a}\otimes(\sigma_{z}^p)_{b}\otimes I_{c}$
			\\\hline
		\end{tabular}\label{tab1}
	\end{center}
\end{table}

In next part shown in Fig. \ref{fig3}, the Toffoli gate operations,
are performed on the spatial DoF of three-photon, by utilizing two cavity-NV systems and five block$_{2}$ to finish
system $abc$, without changing the polarized states of
three photons. Suppose that the states of the three photons are the same as ones described in Eq. (\ref{eq5}), and meanwhile the states of  NV$_{3}$ and  NV$_{4}$ are $|+1\rangle_{3}$ and $|-1\rangle_{4}$, respectively.

First, after the Hadamard operations H$_{e}$ are performed on the two electron-spin states in NV$_{3}$  and NV$_{4}$,
let the photons $a$ and
$b$ simultaneously enter into path $|0\rangle_{a}^s$ (or $|1\rangle_{a}^s$) and $|0\rangle_{b}^s$ (or $|1\rangle_{b}^s$) from opposite directions, respectively. In detail, photon $a$ from the spatial path  $|1\rangle_{a}^s$  interacts with the block$_{2}$-NV$_{4}$, meanwhile, photon $b$ from the spatial path  $|1\rangle_{b}^s$ interacts with the block$_{2}$-NV$_{3}$. Both photon $a$ from the spatial path   $|0\rangle_{a}^s$ and photon $b$ from the spatial path $|0\rangle_{b}^s$  pass through a VBS$_{1}$ and a DL in sequence.
VBS$_{1}$ is used to adjust the transmission coefficient.  DL is used to make the two paths $|0\rangle_{a}^s$ and $|1\rangle_{a}^s$, and $|0\rangle_{b}^s$ and $|1\rangle_{b}^s$ of two photons $a$ and
$b$ arrive simultaneously.
In the block$_{2}$-NV$_{3}$, the polarized state $|0\rangle^p$ of photon $b$ is reflected by CPBS$_{1}$, then it sequentially passes through H$_{1}$, NV$_{3}$, and H$_{1}$, while the polarized state $|1\rangle^p$ is transmitted to sequentially pass through X$_{1}$, H$_{1}$, NV$_{3}$, H$_{1}$, and X$_{2}$.
After that, the two optical wave packets converge on another CPBS$_{2}$.
So does the photon $a$ in the block$_{2}$-NV$_{4}$.
When all detectors Ds in the block$_{2}$ and VBS$_{1}$s are no response, the state of the whole system is transmitted from $|\psi_{0}\rangle$ into $|\psi_{1}\rangle$. Here,
\begin{eqnarray}\label{eq11}
	|\psi_{0}\rangle&=&	|\phi\rangle_{a}^{p}\otimes|\phi\rangle_{a}^{s}\otimes|\phi\rangle_{b}^{p}\otimes|\phi\rangle_{b}^{s}\otimes|\phi\rangle_{c}^{p}\otimes|\phi\rangle_{c}^{s}\nonumber\\
	&&\otimes|+1\rangle_{3}\otimes|-1\rangle_{4},\nonumber\\
	|\psi_{1}\rangle  &=  & \frac{r_{1}-r_{0}}{2\sqrt{2}}  [|+1\rangle_{4}(\delta_{1}|0\rangle_{a}+\delta_{2}|1\rangle_{a})\nonumber\\
	&&+|-1\rangle_{4}(-\delta_{1}|0\rangle_{a}+\delta_{2}|1\rangle_{a})]\nonumber\\
	&&\frac{r_{1}-r_{0}}{2\sqrt{2}}[|+1\rangle_{3}(\epsilon_{1}|0\rangle_{b}^s+\epsilon_{2}|1\rangle_{b}^s)\nonumber\\
	&&+|-1\rangle_{3}(  \epsilon_{1}|0\rangle_{b}^s-\epsilon_{2}|1\rangle_{b}^s  )]\nonumber\\
	&&\otimes  |\phi\rangle_{a}^{p}|\phi\rangle_{b}^{p} |\phi\rangle_{c}^{p} |\phi\rangle_{c}^{s}.
\end{eqnarray}
From Eq. (\ref{eq11}),  the function of block$_{2}$-NV$_{j}$ ($j$ = 3, 4) is equal to performing the CPF gate operation on the spatial
state $|0\rangle^{s}$ of photon $a$ or
$b$ when the electron-spin state $|-1\rangle$ in NV$_{i}$ ($j$ = 3, 4) center, without affecting its polarized state.
Similarly, when the detector D in the block$_{2}$ on the spatial mode  $|1\rangle_{a}^s$ ($|1\rangle_{b}^s$) or  VBS$_{1}$ on the spatial mode $|0\rangle_{a}^s$ ($|0\rangle_{b}^s$) responses,
the errors caused by the imperfect photon scattering are heralded by the D.
That is,  the
block$_{2}$  also has an error-detected function.
After the photons $a$ from the spatial path $|1\rangle_{a}^s$ and
$b$ from the spatial path  $|1\rangle_{b}^s$ interact with  block$_{2}$-NV$_{4}$ and block$_{2}$-NV$_{3}$, respectively, the  two electron-spin states in NV$_{3}$ and NV$_{4}$ centers perform
the Hadamard operations H$_{e}$ again. The quautum state $|\psi_{1}\rangle$ is changed into $|\psi_{1'}\rangle$
\begin{eqnarray}   \label{eq12}
	|\psi_{1'}\rangle  &= & (\frac{r_{1}-r_{0}}{2})^2    (\delta_{1}|0\rangle_{a}^s|-1\rangle_{4}     +   \delta_{2}|1\rangle_{a}^s|+1\rangle_{4}) \nonumber\\
	&& \otimes    (\epsilon_{1}|0\rangle_{b}^s|+1\rangle_{3}   +     \epsilon_{2}|1\rangle_{b}^s|-1\rangle_{3}) \nonumber\\
	&&    \otimes  |\phi\rangle_{a}^{p}     |\phi\rangle_{b}^{p}    |\phi\rangle_{c}^{p}      |\phi\rangle_{c}^{s}.
\end{eqnarray}

Second, the photon $c$  is put the quantum circuit.
The photon $c$ from the spatial state $|0\rangle_{c}^s$ passes through a VBS$_{1}$ and a DL in sequence.
After
the photon $c$ from the spatial path  $|1\rangle_{c}^s$ passes through BS$_{1}$,  it completes the transformation, i.e., $|1\rangle_{c}^s\leftrightarrow(|0\rangle_{d}^s+|1\rangle_{d}^s)/\sqrt{2}$. Subsequently, the photon $c$ from the spatial path $|1\rangle_{d}^s$ passes through block$_{2}$-NV$_{3}$, BS$_{2}$, block$_{2}$-NV$_{4}$, BS$_{3}$, DL, VBS$_{1}$ in sequence,  while the photon $c$ from the spatial path $|0\rangle_{d}^s$ passes through VBS$_{1}$, DL, BS$_{2}$, VBS$_{1}$, DL, BS$_{3}$,  block$_{2}$-NV$_{3}$ in sequence. Then
the photon $c$ from the spatial path $|0\rangle_{d}^s$ and $|1\rangle_{d}^s$  converge at BS$_{4}$.  Here,  BS$_{k}$(k=2,3) is used to completes the Hadamard operation on the spatial DoF of the photon $c$, i.e., $|0\rangle_{d}^s\leftrightarrow(|0\rangle_{d}^s+|1\rangle_{d}^s)/\sqrt{2}$, and $|1\rangle_{d}^s\leftrightarrow(|0\rangle_{d}^s-|1\rangle_{d}^s)/\sqrt{2}$.
DLs are used to make two spatial paths $|0\rangle_{d}^s$ and $|1\rangle_{d}^s$ of the photon $c$ converge at simultaneously BS$_{2}$, BS$_{3}$,  and BS$_{4}$, respectively.
DLs from the spatial paths $|0\rangle_{a}^s$,  and $|0\rangle_{b}^s$,  $|0\rangle_{c}^s$ make the photons $a$, $b$, and $c$ arrive simultaneously, respectively.
When all detectors Ds in three block$_{2}$s , VBS$_{1}$s and VBS$_{2}$s do not response, the above operations change the quautum state $|\psi_{1'}\rangle$ into
\begin{eqnarray}   \label{eq13}
	\!\!\!\!\!\!\!\!\!	|\psi_{2}\rangle&=&(\frac{r_{1}-r_{0}}{2})^5[(\delta_{1}\epsilon_{1}|0\rangle_{a}^s|0\rangle_{b}^s+\delta_{1}\epsilon_{2}|0\rangle_{a}^s|1\rangle_{b}^s\nonumber\\
	&&+\delta_{2}\epsilon_{1}|1\rangle_{a}^s|0\rangle_{b}^s)(\zeta_{1}|0\rangle_{c}^s+\zeta_{2}|1\rangle_{c}^s)\nonumber\\
	&&+(\delta_{2}\epsilon_{2}|1\rangle_{a}^s|1\rangle_{b}^s)(\zeta_{1}|0\rangle_{c}^s-\zeta_{2}|1\rangle_{c}^s)]\nonumber\\
	&&\otimes|\phi\rangle_{a}^{p}|\phi\rangle_{b}^{p}|\phi\rangle_{c}^{p}.
\end{eqnarray}
Third, we apply Hadamard operations H$_{e}$ on the two electron-spin states in NV$_{3}$ and NV$_{4}$ centers, respectively,  and  measure them with the basis  $\left\{|+1\rangle,|-1\rangle\right\}$. If the measurement results of the two electron-spin states are $|+1\rangle_{1}$ and $|+1\rangle_{2}$, The state of the system
is changed from $|\psi_{2}\rangle$  to
\begin{eqnarray} \label{eq14}
	\!\!\!\!\!\!\!\!
	|\psi_{3}\rangle  &=&(\frac{r_{1}-r_{0}}{2})^5[(\delta_{1}\epsilon_{1}|0\rangle_{a}^s|0\rangle_{b}^s+\delta_{1}\epsilon_{2}|0\rangle_{a}^s|1\rangle_{b}^s\nonumber\\
	&&+\delta_{2}\epsilon_{1}|1\rangle_{a}^s|0\rangle_{b}^s)(\zeta_{1}|0\rangle_{c}^s+\zeta_{2}|1\rangle_{c}^s)\nonumber\\
	&&+ \delta_{2}\epsilon_{2}|1\rangle_{a}^s|1\rangle_{b}^s(\zeta_{1}|0\rangle_{c}^s-\zeta_{2}|1\rangle_{c}^s)]\nonumber\\
	&&\otimes  |\phi\rangle_{a}^{p}     |\phi\rangle_{b}^{p}     |\phi\rangle_{c}^{p} ,
\end{eqnarray}
which is the result of the C$^2$PF gate on the spatial DoF of the three-photon
system without affecting its polarized state. That is, when the two electron-spin states in NV$_{3}$ and NV$_{4}$, and the
polarized states of two photons $ab$ are $|+1\rangle_{3}|+1\rangle_{4}$ and $|1\rangle_{a}^s|1\rangle_{b}^s$, the CPF gate operation on the spatial
state of photon $c$.
However, the measurement results of the two-electron-spin states are other possible outcomes, the corresponding feed-forward single-qubit operations are shown in Table \ref{tab2}. Here, phase shifter $e^{i\pi}$ ,which completes the transformation $|0\rangle^{p}\rightarrow -|0\rangle^{p}$ and $|1\rangle^{p}\rightarrow -|1\rangle^{p}$, is performed on spatial state, and $I$ means keeping its original state. Therefore, the success probability of a C$^2$PF gate on the spatial DoF of a three-photon
system is 100\% in principle.
Finally, before and after the C$^2$PF gate,  Hadamard operation is performed on the spatial DoF of the target qubit $c$, i.e., $U^s_{Toffoli} =
(H^s \otimes I_{4})U^s_{C^2PF}(I_{4} \otimes H^s)$, the Toffoli gate on the spatial DoF of the three-photon system is finished without impacting on the polarized DoF of the three-photon system. Here,
\begin{eqnarray} \label{eq15}
	\!\!\!\!\!\!\!\!|\psi_{4}\rangle  &=&(\frac{r_{1}-r_{0}}{2})^5[(\delta_{1}\epsilon_{1}|0\rangle_{a}^s|0\rangle_{b}^s+\delta_{1}\epsilon_{2}|0\rangle_{a}^s|1\rangle_{b}^s\nonumber\\
	&&+\delta_{2}\epsilon_{1}|1\rangle_{a}^s|0\rangle_{b}^s)(\zeta_{1}|0\rangle_{c}^s+\zeta_{2}|1\rangle_{c}^s)\nonumber\\
	&&+ \delta_{2}\epsilon_{2}|1\rangle_{a}^s|1\rangle_{b}^s(\zeta_{2}|0\rangle_{c}^s+\zeta_{1}|1\rangle_{c}^s)]\nonumber\\
	&&\otimes  |\phi\rangle_{a}^{p}     |\phi\rangle_{b}^{p}     |\phi\rangle_{c}^{p}  .
\end{eqnarray}

Finally, we can combine the above two parts in Fig. \ref{fig2} and \ref{fig3} to construct an  error-detected hyper-parallel Toffoli gate with state-selective reflection, as the Toffoli gates of the three-photon system in
the polarized and spatial DoFs
are independent of each other.
The hybrid quantum CPF gate operations are the key elements of
this hyper-Toffoli gate.
The above two Tables \ref{tab1} and \ref{tab2} are also valid when the four electron spins of the NV-cavity systems are in other states.
Thus, our error-detected hyper-parallel Toffoli gate is very flexible.	 Here,
\begin{eqnarray} \label{eq16}
	\!\!\!\!\!\!\!\!\!
	|\Phi\rangle &=&(\frac{r_{1}-r_{0}}{2})^5[(\alpha_{1}\beta_{1}|0\rangle_{a}^p|0\rangle_{b}^p+\alpha_{1}\beta_{2}|0\rangle_{a}^p|1\rangle_{b}^p\nonumber\\
	&&+\alpha_{2}\beta_{1}|1\rangle_{a}^p|0\rangle_{b}^p)(\gamma_{1}|0\rangle_{c}^p+\gamma_{2}|1\rangle_{c}^p)\nonumber\\
	&&+(\alpha_{2}\beta_{2}|1\rangle_{a}^p|1\rangle_{b}^p)(\gamma_{1}|1\rangle_{c}^p+\gamma_{2}|0\rangle_{c}^p)]\nonumber\\ &&\otimes(\frac{r_{1}-r_{0}}{2})^5[(\delta_{1}\epsilon_{1}|0\rangle_{a}^s|0\rangle_{b}^s+\delta_{1}\epsilon_{2}|0\rangle_{a}^s|1\rangle_{b}^s\nonumber\\
	&&+\delta_{2}\epsilon_{1}|1\rangle_{a}^s|0\rangle_{b}^s) (\zeta_{1}|0\rangle_{c}^s+\zeta_{2}|1\rangle_{c}^s)\nonumber\\
	&&+ (\delta_{2}\epsilon_{2}|1\rangle_{a}^s|1\rangle_{b}^s)(\zeta_{1}|1\rangle_{c}^s+\zeta_{2}|0\rangle_{c}^s)].
\end{eqnarray}
As the incomplete and imperfect cavity-NV-center interactions are transformed into the detectable failure rather than infidelity
based on
the novel
heralding mechanism of detectors in the
block$_{1}$ and
block$_{2}$,
our hyper-Toffoli gate of the three-photon system is error-detected.
The efficiencies of the two blocks can be further improved by repeating the operation processes when the detectors in blocks are clicked.
Now, we have
obtained the result of the error-detected hyper-parallel Toffoli gate with state-selective reflection of one-sided cavity-NV center
system. That is, when the polarized state of two photons $ab$ are $|1\rangle_{a}^p|1\rangle_{b}^p$,
the polarized states of the photon $c$ is flipped; when
the spatial state of two photons $ab$ are $|1\rangle_{a}^s|1\rangle_{b}^s$, the spatial states
of the photon $c$ is flipped. No operation is performed
on the photon $c$ when the polarized state of two photons $ab$
are $|0\rangle_{a}^p|0\rangle_{b}^p$ ($|0\rangle_{a}^p|1\rangle_{b}^p$ or $|1\rangle_{a}^p|0\rangle_{b}^p$) and the spatial state of two photons $ab$ are $|0\rangle_{a}^s|0\rangle_{b}^s$ ($|0\rangle_{a}^s|1\rangle_{b}^s$ or $|1\rangle_{a}^s|0\rangle_{b}^s$).

\begin{table} [ht]
	\tabcolsep 0pt \caption{The measurement results of the two-electron-spin states in NV$_{3}$ and NV$_{4}$, and the corresponding feed-forward single-qubit operations on polarized DoF of three photons.} 
	\begin{center}
		\begin{tabular}{cc}
			\hline
			measurement results \ \ \ \
			& single-qubit operations \\\hline
			$|+1\rangle_{3}|+1\rangle_{4}$\ \
			&$I_{a}\otimes I_{b} \otimes I_{c}$\\
			$|+1\rangle_{3}|-1\rangle_{4} $\ \
			& $e^{i\pi}|0\rangle_{a}^s \otimes I_{b} \otimes I_{c}$\\
			$|-1\rangle_{3}|+1\rangle_{4}$\ \
			& $I_{a}\otimes  e^{i\pi}|1\rangle_{b}^s\otimes I_{c}$\\
			\ $|-1\rangle_{3}|-1\rangle_{4}$ \ \
			& $e^{i\pi}|0\rangle_{a}^s\otimes e^{i\pi}|1\rangle_{b}^s \otimes I_{c}$   	
			\\\hline
		\end{tabular}\label{tab2}
	\end{center}
\end{table}

\section{Discussion and conclusion}

In this work, we have designed an error-detected quantum circuit for implementing the hyper-parallel Toffoli gate on a three-photon system in both the polarization and spatial DoFs, assisted by state-selective reflection of one-sided cavity-NV-center system. The hyper-Toffoli gate can perform double Toffoli gate
operations on only one three-photon system in two DoFs
with four one-sided cavity-NV center systems, which can save the
resource depletion and decrease the photonic dispassion noise
in QIP. The interaction of the photons
is accomplished by the photon-electron-spin interaction in the
NV center, which can be greatly improve by coupled to an optical cavity, fiber-based microcavity or microring resonator either in strong coupling regime and in weak coupling regime in experiment \cite{Room-temperature2006,Gigahertz2009,Deterministric2010,Experimental2015,Coupling2013,Integrated2012,Strain2014}.
The identically optical transition of the four uncorrelated NV centers,
can be achieved by applying controlled external electric fields \cite{Observation2009,Electrical2011}.
Our scheme nearly immunes to spectral diffusion and charge fluctuation because of the narrow linewidth
of the state $|A_{2}\rangle$ \cite{Entangled2015}. Many techniques have been explored to reduce and eliminate the spectral-diffusion influence \cite{High-filelity2011,Quantum2007,Room2010,Optically2013}.

\begin{figure}[htp]
	\subfigure[]{
		\begin{minipage}[b]{0.9\linewidth}
			\includegraphics[width=0.9\linewidth]{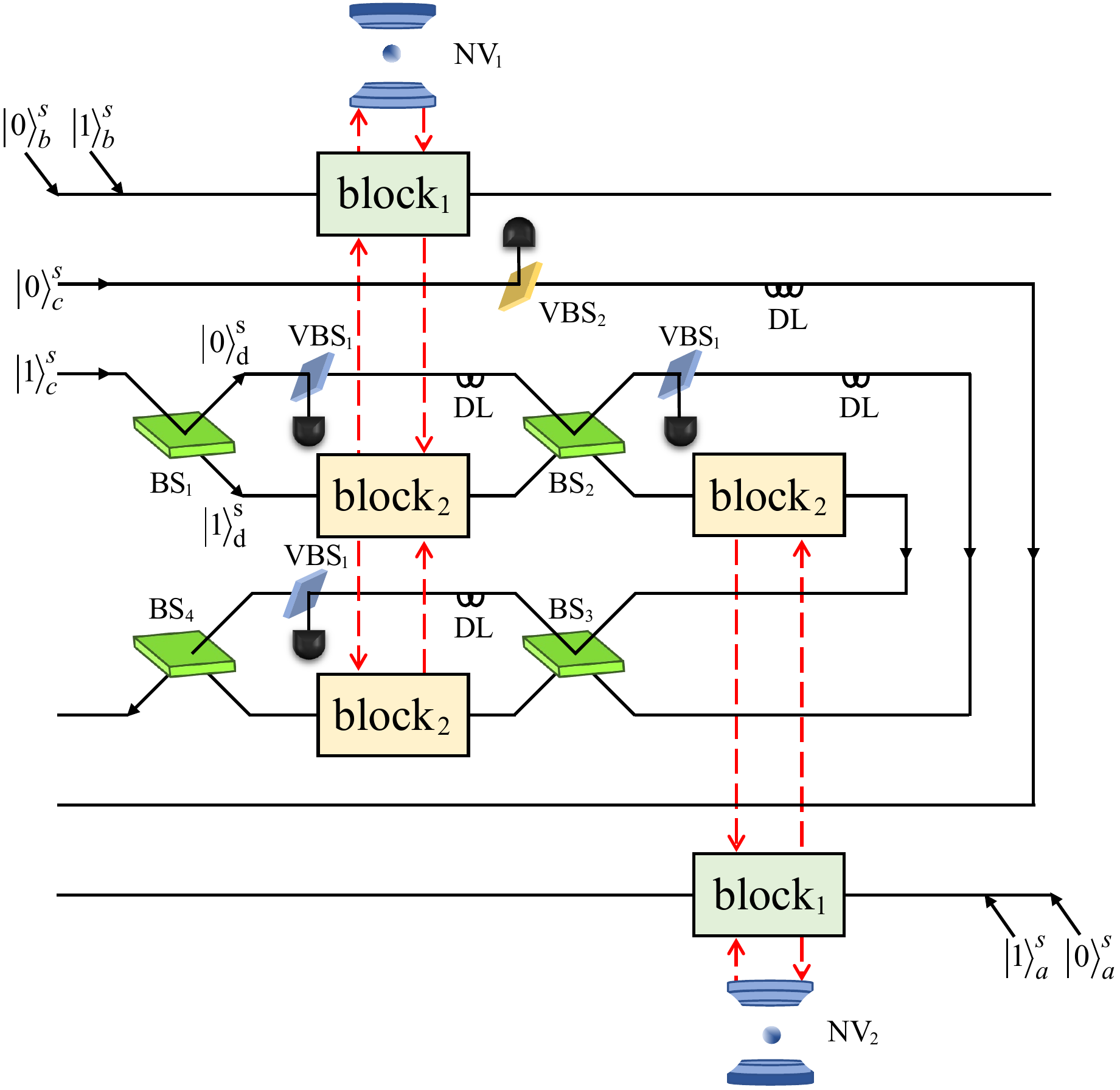}
			\label{fig4a}
	\end{minipage}}	
	\subfigure[]{
		\begin{minipage}[t]{0.9\linewidth}
			\includegraphics[width=0.9\linewidth]{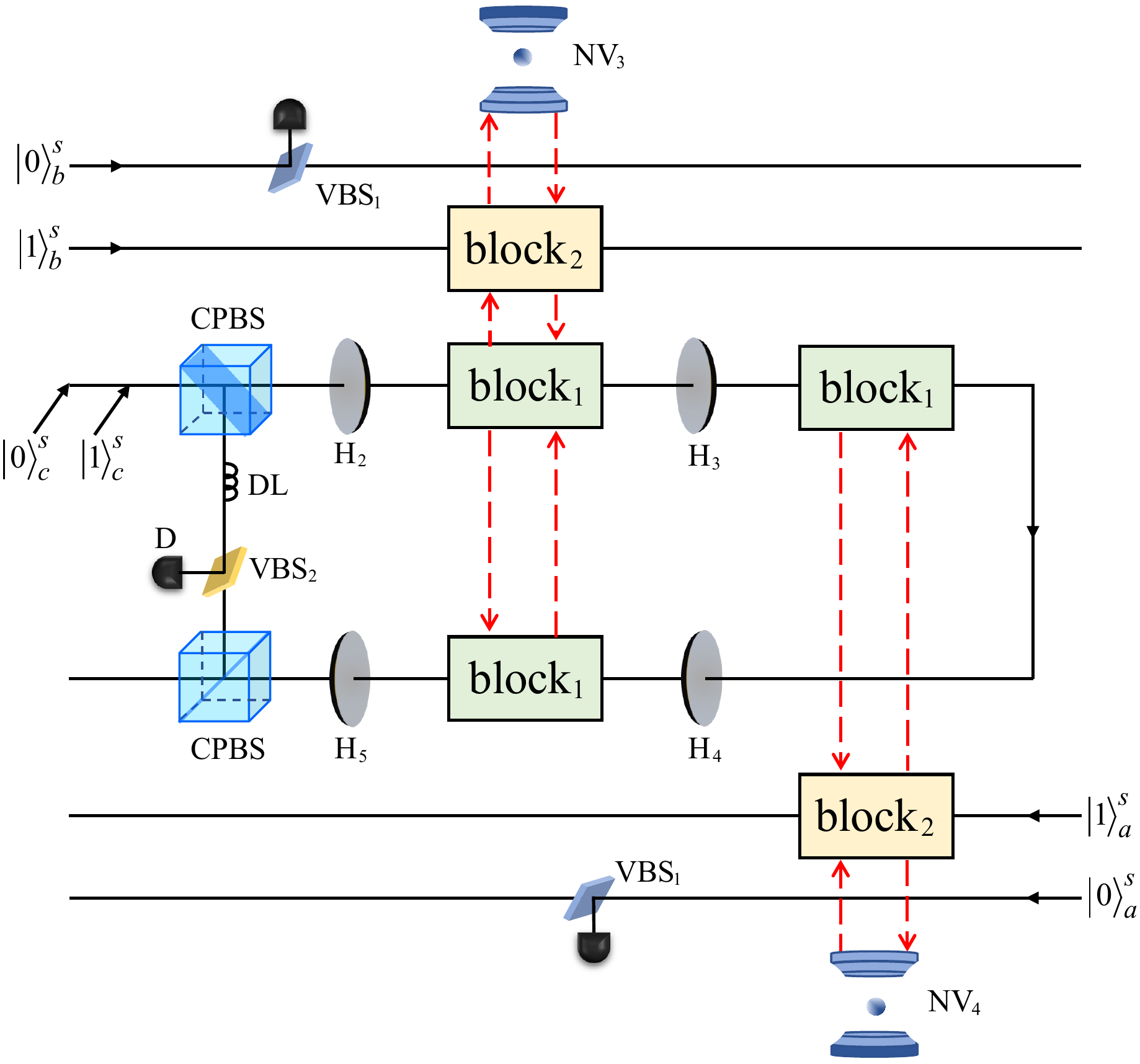}
			\label{fig4b}
	\end{minipage}}
	\caption{Schematic diagram of a  hybrid hyper-Toffoli gate. }\label{fig4}
\end{figure}

We can also construct three types of flexible hybrid hyper-Toffoli gate assisted by the state-selective reflection of one-sided cavity-NV-center system.
The quantum circuit of the hybrid hyper-Toffoli gates in Fig. \ref{fig4} is similar to the one of
the hyper-Toffoli gate in Fig. \ref{fig2} and  Fig. \ref{fig3}, by replacing  H and block$_1$ with BS and block$_2$, respectively.
As shown in Fig. \ref{fig4}, we construct a hybrid hyper-Toffoli gate, that is, the polarized (spatial) states of two photons $a$ and $b$ to control the spatial (polarized) state of photon $c$, respectively.
In detail,  if all Ds can not click of Fig. \ref{fig4a},  the quantum circuit finishes
the polarized states of two photons $a$ and $b$ to control the spatial state of photon $c$. Here,
\begin{eqnarray} \label{eq17}
	|\phi_{h}\rangle &=&(\frac{r_{1}-r_{0}}{2})^5[(\alpha_{1}\beta_{1}|0\rangle_{a}^p|0\rangle_{b}^p+\alpha_{1}\beta_{2}|0\rangle_{a}^p|1\rangle_{b}^p\nonumber\\
	&&+\alpha_{2}\beta_{1}|1\rangle_{a}^p|0\rangle_{b}^p)(\zeta_{1}|0\rangle_{c}^s+\zeta_{2}|1\rangle_{c}^s)\nonumber\\
	&&+(\alpha_{2}\beta_{2}|1\rangle_{a}^p|1\rangle_{b}^p)(\zeta_{1}|1\rangle_{c}^s+\zeta_{2}|0\rangle_{c}^s)]\nonumber\\
	&&\otimes|\phi\rangle_{a}^{s}|\phi\rangle_{b}^{s}|\phi\rangle_{c}^{p}.
\end{eqnarray}
If all Ds in Fig. \ref{fig4b} can not click, the quantum circuit completes that
the spatial states of two photons $a$ and $b$ control the polarized state of photon $c$. Here,
\begin{eqnarray} \label{eq18}
	|\phi_{t}\rangle &=&(\frac{r_{1}-r_{0}}{2})^5[(\delta_{1}\epsilon_{1}|0\rangle_{a}^s|0\rangle_{b}^s+\delta_{1}\epsilon_{2}|0\rangle_{a}^s|1\rangle_{b}^s\nonumber\\
	&&+\delta_{2}\epsilon_{1}|1\rangle_{a}^s|0\rangle_{b}^s)(\gamma_{1}|0\rangle_{c}^p+\gamma_{2}|1\rangle_{c}^p)\nonumber\\
	&&+ (\delta_{2}\epsilon_{2}|1\rangle_{a}^s|1\rangle_{b}^s)(\gamma_{1}|1\rangle_{c}^p+\gamma_{2}|0\rangle_{c}^p)]\nonumber\\
	&&\otimes|\phi\rangle_{a}^{p}|\phi\rangle_{b}^{p}|\phi\rangle_{c}^{s}.
\end{eqnarray}
Obviously, we can unite the two parts to achieve a hybrid hyper-paralleled Toffoli gate with state-selective reflection. Here,
\begin{eqnarray} \label{eq19}
	\!\!\!\!\!\!\!\!\!
	|\Phi_{1}\rangle &=&(\frac{r_{1}-r_{0}}{2})^5[(\alpha_{1}\beta_{1}|0\rangle_{a}^p|0\rangle_{b}^p+\alpha_{1}\beta_{2}|0\rangle_{a}^p|1\rangle_{b}^p\nonumber\\
	&&+\alpha_{2}\beta_{1}|1\rangle_{a}^p|0\rangle_{b}^p)(\zeta_{1}|0\rangle_{c}^s+\zeta_{2}|1\rangle_{c}^s)\nonumber\\
	&&+(\alpha_{2}\beta_{2}|1\rangle_{a}^p|1\rangle_{b}^p)(\zeta_{1}|1\rangle_{c}^s+\zeta_{2}|0\rangle_{c}^s)]\nonumber\\ &&\otimes(\frac{r_{1}-r_{0}}{2})^5[(\delta_{1}\epsilon_{1}|0\rangle_{a}^s|0\rangle_{b}^s+\delta_{1}\epsilon_{2}|0\rangle_{a}^s|1\rangle_{b}^s\nonumber\\
	&&+\delta_{2}\epsilon_{1}|1\rangle_{a}^s|0\rangle_{b}^s)(\gamma_{1}|0\rangle_{c}^p+\gamma_{2}|1\rangle_{c}^p)\nonumber\\
	&&+ (\delta_{2}\epsilon_{2}|1\rangle_{a}^s|1\rangle_{b}^s)(\gamma_{1}|1\rangle_{c}^p+\gamma_{2}|0\rangle_{c}^p)].
\end{eqnarray}
Similarly, we can also construct other types of flexible hybrid hyper-Toffoli gate assisted by the state-selective reflection of one-sided cavity-NV-center system. For example, the polarized (spatial) states of photon $a$ and the spatial (polarized) states of photon $b$ control the spatial (polarized) state of photon $c$, respectively. Here,
\begin{eqnarray} \label{eq20}
	|\Phi_{2}\rangle &=&(\frac{r_{1}-r_{0}}{2})^5[(\alpha_{1}\epsilon_{1}|0\rangle_{a}^p|0\rangle_{b}^s+\alpha_{1}\epsilon_{2}|0\rangle_{a}^p|1\rangle_{b}^s\nonumber\\
	&&+\alpha_{2}\epsilon_{1}|1\rangle_{a}^p|0\rangle_{b}^s)(\zeta_{1}|0\rangle_{c}^s+\zeta_{2}|1\rangle_{c}^s)\nonumber\\
	&&+(\alpha_{2}\epsilon_{2}|1\rangle_{a}^p|1\rangle_{b}^s)(\zeta_{1}|1\rangle_{c}^s+\zeta_{2}|0\rangle_{c}^s)]\nonumber\\ &&\otimes(\frac{r_{1}-r_{0}}{2})^5[(\delta_{1}\beta_{1}|0\rangle_{a}^s|0\rangle_{b}^p+\delta_{1}\beta_{2}|0\rangle_{a}^s|1\rangle_{b}^p\nonumber\\
	&&+\delta_{2}\beta_{1}|1\rangle_{a}^s|0\rangle_{b}^p)(\gamma_{1}|0\rangle_{c}^p+\gamma_{2}|1\rangle_{c}^p)\nonumber\\
	&&+ (\delta_{2}\beta_{2}|1\rangle_{a}^s|1\rangle_{b}^p)(\gamma_{1}|1\rangle_{c}^p+\gamma_{2}|0\rangle_{c}^p)],\nonumber\\
	|\Phi_{3}\rangle &=&(\frac{r_{1}-r_{0}}{2})^5[(\alpha_{1}\epsilon_{1}|0\rangle_{a}^p|0\rangle_{b}^s+\alpha_{1}\epsilon_{2}|0\rangle_{a}^p|1\rangle_{b}^s\nonumber\\
	&&+\alpha_{2}\epsilon_{1}|1\rangle_{a}^p|0\rangle_{b}^s)(\gamma_{1}|0\rangle_{c}^p+\gamma_{2}|1\rangle_{c}^p)\nonumber\\
	&&+(\alpha_{2}\beta_{2}|1\rangle_{a}^p|1\rangle_{b}^p)(\gamma_{1}|1\rangle_{c}^p+\gamma_{2}|0\rangle_{c}^p)]\nonumber\\ &&\otimes(\frac{r_{1}-r_{0}}{2})^5[(\delta_{1}\beta_{1}|0\rangle_{a}^s|0\rangle_{b}^p+\delta_{1}\beta_{2}|0\rangle_{a}^s|1\rangle_{b}^p\nonumber\\
	&&+\delta_{2}\beta_{1}|1\rangle_{a}^s|0\rangle_{b}^p)(\zeta_{1}|0\rangle_{c}^s+\zeta_{2}|1\rangle_{c}^s)\nonumber\\
	&&+ (\delta_{2}\epsilon_{2}|1\rangle_{a}^s|1\rangle_{b}^s)(\zeta_{1}|1\rangle_{c}^s+\zeta_{2}|0\rangle_{c}^s)].
\end{eqnarray}

\begin{table} [ht]
	\tabcolsep 0pt \caption{The efficiency $\eta_T$ of the hyper-Toffoli gates  and the probability $\eta_D$ of
		D triggered in the
		block$_{1}$ or block$_{2}$  vs the ratio of
		$g/\kappa$ when $\omega_{c}=\omega_{p}=\omega_{c}$ at  $\gamma=0.01\kappa$. } 
	\begin{center}
		\begin{tabular}{c|ccc}
			\hline
			$ $ \qquad
			& \ $g/\kappa = 0.5$ \ \ \
			& \ $g/\kappa = 1.5$ \ \ \
			& \ $g/\kappa = 2.4$  \\\hline
			\ $\eta_T$\
			& 90.53$\%$
			& 98.90$\%$
			& 99.57$\%$  \\
			\ $\eta_D$\
			& 0.01$\%$
			& $0$
			& $0$
			\\\hline
		\end{tabular}\label{tab3}
	\end{center}
\end{table}

\begin{figure}[htp!]
	\centering
	\includegraphics[width=0.8\linewidth]{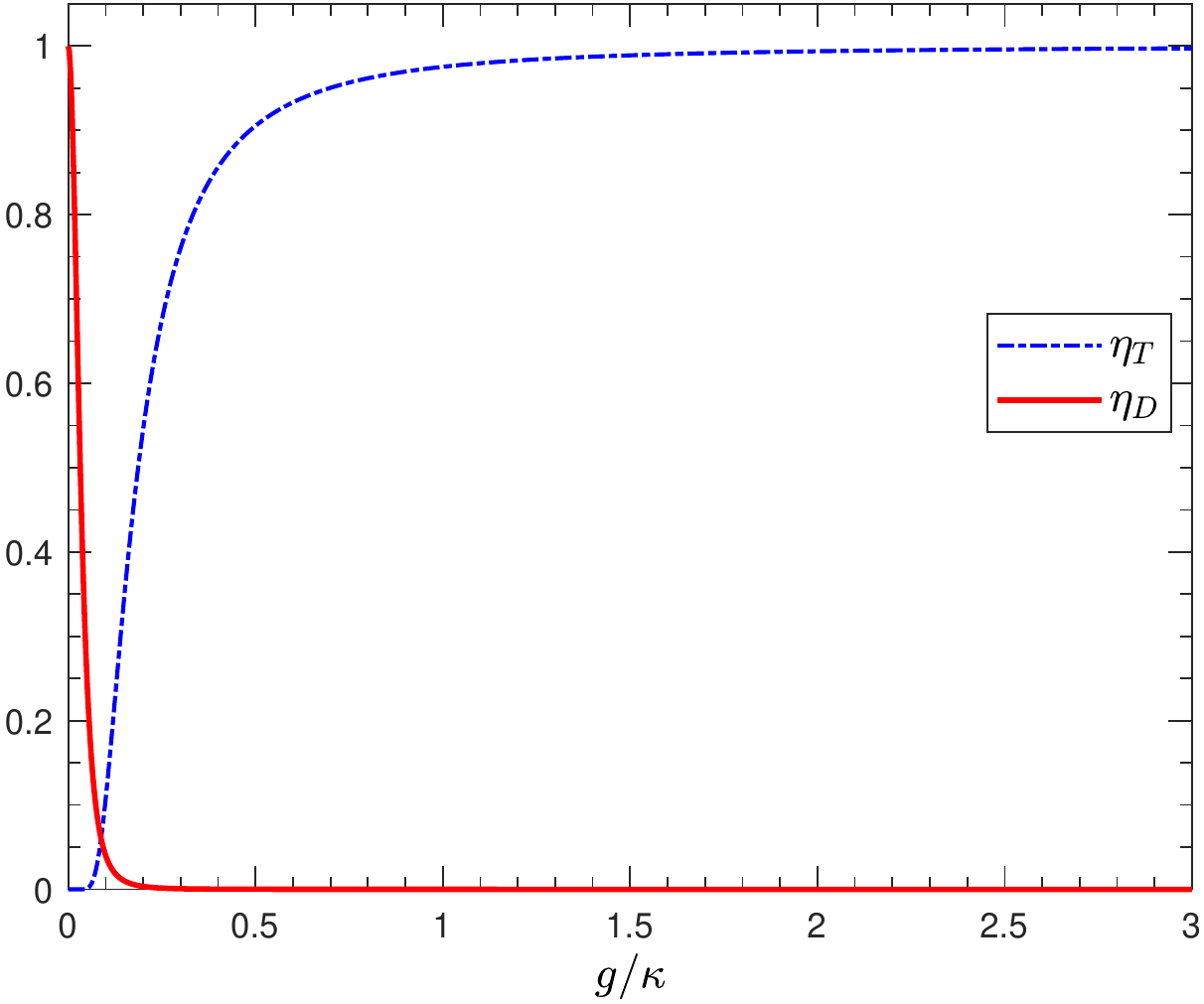}
	\caption{ The efficiency $\eta_T$ of the hyper-Toffoli gates  and the probability $\eta_D$ of
		D triggered in the
		block$_{1}$  vs the ratio of
		$g/\kappa$ when $\omega_{c}=\omega_{p}=\omega_{c}$ at  $\gamma=0.01\kappa$. }
	\label{fig5}
\end{figure}

In the above protocols, we have taken into account practical experiment environment, including the cavity leakage rate and the coupling strength.
The probability $\eta_D^p$ of
D triggered in the
block$_{1}$ and the one $\eta_D^s$ in the
block$_{2}$ is equal to $\eta_D^p=\eta_D^s=\eta_D=(\frac{r_{1}+r_{0}}{2})^2$, in Fig. \ref{fig5}, which are related to the coupling strength $g$, cavity decay and leakage rate $\kappa, \gamma$ at $\gamma=0.01\kappa$. Obviously,
the probability $\eta_D$  shown in Table \ref{tab3}
exists zero values even in the weak coupling rate, which results from the destructive interference.
Therefore, the fidelity of the hyper-Toffoli gates is unity in principle without strong couple limitation between photon and NV center.
These positive characteristics may make our projects conducive to increase the information processing capability and decrease the complexity of large-scale integration.
The efficiency of the hyper-Toffoli gates is  $\eta_T=(\frac{r_{1}-r_{0}}{2})^{10}$, the ratio of the number of output photons to the input.
The relation between $\eta_T$ and $g/\kappa$ is shown in Fig. \ref{fig5} and the partial detailed data is shown in Table \ref{tab3}.
One can see that our schemes can work not only in strong coupling rate but also in the weak coupling rate.
To achieve a higher efficiency $\eta_T$, $\gamma/\kappa$ should be as small as possible, which can be realized by improving the design and manufacturing process of cavity.
Besides, the efficiency of our scheme can also reduced from non-ideal single-photon sources, imperfect linear-optical
elements (BSs and CPBSs), and invalid dark detectors, which
will be improved with the further development of the current technology.

In summary, we have proposed an error-detected hyper-Toffoli gate for the three-photon system based on the interface between polarized photon and cavity-NV-center system. This hyper-Toffoli gate is equivalent to perform double Toffoli gate operations simultaneously on both the polarization and spatial DoFs of the three-photon system with a low decoherence, short operation time, and less quantum resources required, in compared with those on two independent three-photon systems in one DoF only.
Based on the heralding mechanism of detectors, the imperfect interactions between the cavity-NV center and the photon can be translated into judging the response status of the detectors, so the fidelity of our proposal is near unity. The efficiency of our optimal protocol for the hyper-Toffoli gate can be improved further by repeating the operation
processes when the detectors are clicked.
We can also
construct three types of hybrid hyper-Toffoli gate assisted by the state-selective reflection of one-sided cavity-NV-center system, as the quantum circuit of the hyper-Toffoli gate is flexible and adjustable.
Besides,
the modular circuits effectively reduce the complexity of Toffoli gate, and beyond that, two  photons  enter the systems simultaneously from opposite directions can effectively reduce the overall running time.
Moreover,
the evaluation of gate performance with achieved experiment parameters, even in the weak coupling, shows that it is feasible with current experimental technology and provides a promising building block for quantum compute.

\section*{Acknowledgments}

This work was supported in part by the Natural Science Foundation of China under Contract 61901420;	
in part by the Shanxi Province Science Foundation for Youths under Contract 201901D211235;
in part by the Scientific and Technological Innovation Programs
of Higher Education Institutions in Shanxi under Contract 2019L0507;
in part by the Shanxi "1331 Project" Key Subjects Construction.

.\end{document}